\documentclass[aps,prl,twocolumn,amsmath,amssymb,nofootinbib,superscriptaddress]{revtex4}

\usepackage[dvips]{graphicx}
\usepackage{mathrsfs}

\begin{document}

\bibliographystyle{apsrev}

\title{Are quantum walks the saviour of optical quantum computing?}

\author{Peter P. Rohde}
\email[]{dr.rohde@gmail.com}
\homepage{http://peterrohde.wordpress.com}
\affiliation{Centre for Quantum Computer Technology, University of Queensland, Australia}

\date{\today}

\frenchspacing

\maketitle

Quantum walks have emerged as an interesting candidate for the implementation of quantum information processing protocols \cite{bib:Childs09,bib:Kempe08}. Optical implementations of quantum walks have been demonstrated by various groups \cite{bib:Schreiber10,bib:Broome10,bib:Peruzzo10}, and some have received high-profile coverage including the appearance of Ref. \cite{bib:Peruzzo10} in Science. It is often claimed that quantum walks provide an avenue towards universal quantum computation. In this comment I wish to dispel some misconceptions surrounding the prospects of quantum walks as a route towards universal optical quantum computation.

The universality of quantum walks for quantum computation has been demonstrated in the case of both continuous- \cite{bib:Childs09} and discrete-time \cite{bib:Lovett10} systems. While both of these proofs demonstrate an equivalence between a universal gate set and the quantum walk formalism, the equivalence is not always an efficient. While a universal quantum computer can efficiently simulate a quantum walk, the converse is not necessarily the case. Specifically, in these architectures the required physical resources grow exponentially with the number of qubits being simulated, requiring $2^n$ `quantum wires' to simulate an $n$ qubit circuit. If the quantum wires are to be represented as physical objects, such as optical modes, then an exponential number of optical modes are required. Thus, according to these proofs, if the quantum walk formalism is directly physically implemented, it is as universal for quantum computation as a classical computer is -- it is universal, but the physical resource requirements grow exponentially.

Importantly, I stress that the above criticisms are not directed at the quantum walk formalism per se, but rather at the specific choice of implementation currently being pursued whereby a distinct optical mode is assigned to each vertex in the graph. Indeed, in Ref. \cite{bib:Childs09} it is specifically noted that ``the (quantum walk) construction does not directly give an architecture for a physical device''. The problem with present optical implementations is that the physical architecture being pursued is an inherently non-scalable one.

Nonetheless, some algorithms, primarily graph theoretic in nature, have been described which do not require exponential resources. However these algorithms offer only polynomial speedup compared to the best classical algorithms (unless there are an exponential number of position states). The primary interest in quantum computation is to achieve exponential speedup of algorithms, such as Shor's circuit model algorithm for efficiently factorizing large numbers. No such algorithms have been described for the quantum walk formalism using efficient resources. Indeed, with a single walker (i.e. photon) it is not possible to achieve exponential speedup as the system can be efficiently classically simulated using polynomial space and time resources. For example, the output statistics of a single photon walk can be simulated using coherent light \cite{bib:Schreiber10,bib:RohdeSchreiber10}, which contains only classical interference. It is only with the introduction of higher numbers of walkers that the complexity of the system grows exponentially \cite{bib:RohdeSchreiber10}, non-classical interference takes place, and the possibility of exponential speedup presents itself. While authors are beginning to turn their attention to the multi-walker scenario \cite{bib:Peruzzo10}, no algorithms with exponential speedup have been described and the \emph{efficient} universality of multi-walker walks has not been shown.

With the introduction of higher numbers of walkers decoherence effects become magnified. For example, in an optical implementation, mode-matching, dephasing, photon loss and dark-count requirements become exponentially more stringent. Additionally, as with conventional optical approaches, shifting to the quantum walk formalism does not mitigate the need to simultaneously prepare a large number of photons on-demand, which requires multiple, independent, triggered photon sources, low loss delay lines and fast switching -- technology which is still in its infancy. Thus the quantum walk formalism does not bypass many of the inherent difficulties associated with other optical quantum information processing proposals.

Any scalable quantum computing architecture requires error correction to preserve coherence of the quantum system in the presence of intrinsic or environmental decohering effects \cite{bib:NielsenChuang00}. In the conventional circuit model for quantum computing fault-tolerance theories exists, which prove that provided decoherence can be kept below a fault-tolerance threshold, efficient, large-scale quantum computation can be implemented. In the quantum walk architecture no such theorem has been established. Thus, in the absence of a fault-tolerance theorem it is not apparent that large-scale quantum computation will be possible at all using just quantum walks. Additionally, if a fault-tolerance theorem exists it is not presently clear what the physical resource overheads will be. In the standard circuit model such overheads can be orders of magnitude in size.

In summary, while the exploration of quantum walks is of interest as it demonstrates novel, large-scale interferometric effects which can be used for interesting entanglement generation and for algorithm design, it is not clear that such systems will be of long term value in the direct physical implementation of large scale, and importantly \emph{efficient}, implementation of quantum information processing protocols.

I thank Tim Ralph, Tom Stace, Alessandro Fedrizzi and Andrew White for helpful discussions.

\bibliography{bibliography}

\end{document}